\documentclass[times,5p,twocolumn]{elsarticle}

\usepackage{graphicx}
\usepackage{dcolumn}
\usepackage{bm}
\usepackage[dvips]{color}
\usepackage{epsfig}
\usepackage{xspace}
\usepackage{amssymb,amsmath}
\usepackage{cmap}

\begin{document}

\begin{frontmatter}

\title{Measurement of $D^0$ and $D^+$ meson masses with the KEDR Detector}


\author[binp]{V.V.~Anashin}
\author[binp,nsu]{V.M.~Aulchenko}
\author[binp,nsu]{E.M.~Baldin} 
\author[binp]{A.K.~Barladyan}
\author[binp]{A.Yu.~Barnyakov}
\author[binp]{M.Yu.~Barnyakov}
\author[binp,nsu]{S.E.~Baru}
\author[binp]{I.V.~Bedny} 
\author[binp,nsu]{O.L.~Beloborodova}
\author[binp]{A.E.~Blinov}
\author[binp,nstu]{V.E.~Blinov}
\author[binp]{A.V.~Bobrov}
\author[binp]{V.S.~Bobrovnikov}
\author[binp,nsu]{A.V.~Bogomyagkov}
\author[binp,nsu]{A.E.~Bondar}
\author[binp]{D.V.~Bondarev}
\author[binp]{A.R.~Buzykaev}
\author[binp,nsu]{S.I.~Eidelman}
\author[binp]{Yu.M.~Glukhovchenko}
\author[binp]{V.V.~Gulevich}
\author[binp]{D.V.~Gusev}
\author[binp]{S.E.~Karnaev}
\author[binp]{G.V.~Karpov}
\author[binp]{S.V.~Karpov}
\author[binp,nsu]{T.A.~Kharlamova}
\author[binp]{V.A.~Kiselev}
\author[binp,nsu]{S.A.~Kononov}
\author[binp]{K.Yu.~Kotov}
\author[binp,nsu]{E.A.~Kravchenko}
\author[binp,nsu]{V.F.~Kulikov}
\author[binp,nstu]{G.Ya.~Kurkin}
\author[binp,nsu]{E.A.~Kuper}
\author[binp,nstu]{E.B.~Levichev}
\author[binp]{D.A.~Maksimov}
\author[binp]{V.M.~Malyshev}
\author[binp]{A.L.~Maslennikov}
\author[binp,nsu]{A.S.~Medvedko}
\author[binp,nsu]{O.I.~Meshkov}
\author[binp]{S.I.~Mishnev}
\author[binp,nsu]{I.I.~Morozov}
\author[binp,nsu]{N.Yu.~Muchnoi}
\author[binp]{V.V.~Neufeld}
\author[binp]{S.A.~Nikitin}
\author[binp,nsu]{I.B.~Nikolaev}
\author[binp]{I.N.~Okunev}
\author[binp,nstu]{A.P.~Onuchin}
\author[binp]{S.B.~Oreshkin}
\author[binp,nsu]{I.O.~Orlov}
\author[binp]{A.A.~Osipov}
\author[binp]{S.V.~Peleganchuk}
\author[binp,nstu]{S.G.~Pivovarov}
\author[binp]{P.A.~Piminov}
\author[binp]{V.V.~Petrov}
\author[binp]{A.O.~Poluektov}
\author[binp]{I.N.~Popkov}
\author[binp]{V.G.~Prisekin}
\author[binp]{A.A.~Ruban}
\author[binp]{V.K.~Sandyrev}
\author[binp]{G.A.~Savinov}
\author[binp]{A.G.~Shamov}
\author[binp]{D.N.~Shatilov}
\author[binp,nsu]{B.A.~Shwartz}
\author[binp]{E.A.~Simonov}
\author[binp]{S.V.~Sinyatkin}
\author[binp,nsu]{Yu.I.~Skovpen}
\author[binp]{A.N.~Skrinsky}
\author[binp,nsu]{V.V.~Smaluk}
\author[binp]{A.V.~Sokolov}
\author[binp]{A.M.~Sukharev}
\author[binp,nsu]{E.V.~Starostina}
\author[binp,nsu]{A.A.~Talyshev}
\author[binp]{V.A.~Tayursky}
\author[binp,nsu]{V.I.~Telnov}
\author[binp,nsu]{Yu.A.~Tikhonov}
\author[binp,nsu]{K.Yu.~Todyshev}
\author[binp]{G.M.~Tumaikin}
\author[binp]{Yu.V.~Usov}
\author[binp]{A.I.~Vorobiov}
\author[binp]{A.N.~Yushkov}
\author[binp]{V.N.~Zhilich}
\author[binp,nsu]{V.V.~Zhulanov}
\author[binp,nsu]{A.N.~Zhuravlev}

  \address[binp]{Budker Institute of Nuclear Physics, 11, akademika
  Lavrentieva prospect,  Novosibirsk, 630090, Russia}
  \address[nsu]{Novosibirsk State University, 2, Pirogova street,  Novosibirsk, 630090, Russia}
  \address[nstu]{Novosibirsk State Technical University, 20, Karl Marx
  prospect,  Novosibirsk, 630092, Russia}

\begin{abstract}
  The masses of the neutral and charged $D$ mesons have been
  measured with the KEDR detector at the
  VEPP-4M electron-positron collider:
  \begin{equation*}
    \begin{split}
      M_{D^0}=&1865.30\pm 0.33\pm 0.23 \mbox{ MeV}, \\
      M_{D^+}=&1869.53\pm 0.49\pm 0.20 \mbox{ MeV}.
    \end{split}
  \end{equation*}
\end{abstract}

  \begin{keyword}
    $D$ meson\sep charm\sep mass\sep $X(3872)$\sep $\psi(3770)$
    \PACS 13.20.Fc\sep 13.20.Gd\sep 14.40.Lb
  \end{keyword}
\end{frontmatter}


\renewcommand{\arraystretch}{1.2}


\section{Introduction}

Neutral and charged $D$ mesons are the ground states in the family of
open charm mesons. Measurement of their masses provides a mass scale
for the heavier excited states. In addition, a precise measurement of the
$D^0$ meson mass should help to understand the nature of the narrow
$X(3872)$ state~\cite{Choi:2003ue,Acosta:2003zx,Abazov:2004kp,Aubert:2004ns},
which, according to some models, is a bound state of
$D^0$ and $D^{*0}$ mesons~\cite{Swanson:2003tb} and has a mass very close to
the sum of the $D^0$ and $D^{*0}$ meson masses.
Presently, the world-average $D^0$ mass value~\cite{pdg2008}
($M_{D^0}=1864.84\pm 0.17$~MeV) is dominated by
the CLEO measurement 
$M_{D^0}=1864.847\pm 0.150\mbox{(stat)}\pm 0.095\mbox{(syst)}$~MeV~\cite{cleo},
which uses the decay $D^0\to \phi K_S^0$.
Other $D$ meson mass measurements are much less precise. These
measurements were carried out long ago in the MARK-II experiment at the SPEAR
$e^+e^-$ collider \cite{mark2}, and by the ACCMOR collaboration in a
fixed-target experiment \cite{accmor}.
Both measurements are dominated by the systematic uncertainty, which in the case
of MARK-II is related to beam energy calibration.
In addition,
the mass of the $D^+$ is constrained by the $D^0$ mass and a mass difference
$M_{D^+}-M_{D^0}$ much more precisely than directly measured:
the world-average $D^+$ mass is $M_{D^+}=1869.62\pm 0.20$~MeV,
while the direct measurements yield $M_{D^+}=1869.5\pm 0.5$~MeV.

As both $D^0$ and $D^+$ mass values are based on a single measurement,
the cross-check involving a method different from the one used at CLEO is
essential. This paper describes a measurement which has been performed
with the KEDR detector 
at the VEPP-4M $e^+e^-$ collider using the decay $\psi(3770)\to D\overline{D}$.

\section{Experimental facility}

The electron-positron accelerator complex VEPP-4M~\cite{Anashin:1998sj}
designed for high-energy physics experiments in the
center-of-mass (CM) energy range from 2 to 12 GeV is currently running
in the $\psi$ family region.
The collider consists of two half-rings, an experimental section where
the KEDR detector is installed, and a straight section, which includes an
RF cavity and injection system.
The circumference of the VEPP-4M ring is 366~m. The
luminosity at the $J/\psi$ in an operation mode with 
2 by 2 bunches reaches
$\mathcal{L}=10^{30}$~cm$^{-2}$s$^{-1}$.

Precise measurement of beam energy can be performed at VEPP-4M using
the resonant depolarization method~\cite{Skrinsky:1989ie}. The method is
based on the measurement of the spin precession frequency of the
polarized beam, which depends on its energy. Using resonant
depolarization, the precision of the beam energy measurement reached in the
KEDR experiment is $\simeq$10~keV~\cite{Aulchenko:2003qq}.

The KEDR detector~\cite{Anashin:2002uj} includes
a tracking system consisting of a vertex detector and a drift chamber,
a particle identification (PID) system of aerogel Cherenkov counters and
scintillation time-of-flight counters, and an electromagnetic calorimeter
based on liquid krypton (in the barrel part) and CsI crystals (endcap part). 
The superconducting solenoid provides a longitudinal 
magnetic field of 0.6~T.
A muon system is installed inside the magnet yoke. The detector also
includes a high-resolution tagging system for  studies of two-photon
processes. The online luminosity measurement is performed with sampling
calorimeters which detect photons from the process of single 
brehmsstrahlung.

Charged tracks are reconstructed in the drift chamber (DC) and vertex
detector (VD). DC~\cite{Baru:2002} has a cylindrical shape of 1100 mm length, 
an outer radius of 535 mm and is filled
with pure dimethyl ether. DC cells form seven concentric layers:
four axial layers and three stereo-layers to measure track
coordinates along the beam axis.
The coordinate resolution averaged over drift length is 100 $\mu$m.
VD~\cite{Aulchenko:1989qt} is installed between the vacuum chamber and
DC and increases a solid angle accessible to the tracking system to 98\%.
VD consists of 312 cylindrical drift tubes aligned in 6 layers.
It is filled with an
Ar$+$30\%CO$_2$ gas mixture and has a coordinate resolution of 250~$\mu$m.
The momentum resolution of the tracking system is
$\sigma_p/p=2\%\oplus (4\%\times p$[GeV]$)$.

Scintillation counters of the time-of-flight system (TOF) are
used in a fast charged trigger and for identification of the
charged particles by their flight time. The TOF system consists of
32 plastic scintillation counters in the barrel part and
in each of the endcaps. The flight time resolution is about 350 ps,
which corresponds to $\pi/K$ separation at the level of more than
two standard deviations for momenta up to 650~MeV.

Aerogel Cherenkov counters (ACC)~\cite{instr02atc} are used for particle 
identification in the momentum region not covered by the TOF system
and ionizations measurements in DC. ACC uses aerogel
with the refractive index of 1.05 and 
wavelength shifters for light collection.
This allows one to identify $\pi$ and $K$ mesons in the momentum 
range of 0.6 to 1.5~GeV.
The system design includes 160 counters in the endcap
and barrel parts, each arranged in two layers. During data taking
only one layer of
ACC was installed, and it was not used because of
insufficient efficiency.

The barrel part of the electromagnetic calorimeter is a liquid krypton
ionization detector~\cite{Peleganchuk2009}.
The calorimeter provides an energy
resolution of $3.0$\% at the energy of 1.8 GeV and a spatial resolution
of 0.6--1.0 mm for charged particles and photons. The endcap part
of the calorimeter is based on 1536 CsI(Na) scintillation
crystals~\cite{Aulchenko:1996hi} with an energy
resolution of 3.5\% at 1.8 GeV, and a spatial resolution of 8 mm.

The muon system~\cite{Chilingarov:1988cm} is used to identify muons by their
flight path in the dense medium of the magnetic yoke. It consists of
three layers of streamer tubes with 74\% solid angle coverage,
the total number of channels is 544. The average longitudinal resolution
is 3.5 cm, and the detection efficiency for the most 
of the covered angles is 99\%.

Trigger of the KEDR detector consists of two levels: primary
(PT) and secondary (ST). Both PT and ST operate at the hardware
level. PT uses signals from TOF counters and both calorimeters
as inputs, the typical rate is $5\div 10$ kHz.
ST uses signals from VD, DC and muon system in addition to systems
listed above, and the rate is $50\div150$ Hz.

\section{Measurement method}

\label{method_section}

Measurement of $D$ meson masses is performed using the near-threshold
$e^+e^-\to D\overline{D}$ production with full reconstruction
of one of the $D$ mesons. Neutral $D$ mesons are reconstructed in
the $K^-\pi^+$ final state, charged $D$ mesons are reconstructed in
the $K^-\pi^+\pi^+$ final state (charge-conjugate states are implied
throughout this paper). To increase a data sample, the collider is 
operated at the peak of the $\psi(3770)$ resonance. The production 
cross sections at this energy are
$\sigma(D^0\overline{D}{}^0)=3.66\pm 0.03\pm 0.06$~nb
and $\sigma(D^+D^-)=2.91\pm 0.03\pm 0.05$~nb~\cite{cleo_dd}.

The invariant mass of the $D$ meson can be calculated as
\begin{equation}
    M_{\rm bc}\simeq\sqrt{E_{\rm beam}^2-
               \left(\sum\limits_i \vec{p}_i\right)^2},
\end{equation}
(so-called {\it beam-constrained mass}), where
$E_{\rm beam}$ is the average energy of colliding beams, 
$\vec{p}_i$ are the momenta of the $D$ decay products.
The mass calculated this way is determined more precisely than 
in the case when the $D$ energy is obtained from the energies 
of the decay products.
The precision of $M_D$ measurement in one event is
\begin{equation}
    \sigma^2_{M_{D}}\simeq\sigma^2_{W}/4+\left(\frac{p_D}{M_D}\right)^2\sigma_p^2
                    \simeq\sigma^2_{W}/4+0.02\sigma_p^2\,,
\end{equation}
where $\sigma_W$ is the CM energy spread. The contribution of the
momentum resolution is suppressed significantly due to small $D$
momentum ($p_D\simeq 260$~MeV).

In addition to $M_{\rm bc}$, $D$ mesons are effectively selected by
the CM energy difference
\begin{equation}
    \Delta E=\sum\limits_i \sqrt{M^2_i+p^2_i}-E_{\rm beam}\,,
\end{equation}
where $M_i$ and $p_i$ are the masses and momenta of the $D$ decay products.
The signal events should satisfy a condition $\Delta E\simeq 0$.
In our analysis, we select a relatively wide region of
$M_{\rm bc}$ and $\Delta E$ close to
$M_{\rm bc}\sim M_D$ and $\Delta E\sim 0$ (specifically, 
$M_{\rm bc}>1700$~MeV, $|\Delta E|<300$~MeV); 
then a fit of the
event density is performed with $D$ mass as one of the parameters,
with the background contribution taken into account. 
The background 
in our analysis comes from the random combinations of tracks of  
the continuum process $e^+e^-\to q\bar{q}$ ($q=u,d,s$), from 
other decays of $D$ mesons, and from the signal decays where some 
tracks are picked up from the decay of the other $D$ meson.

While calculating $M_{\rm bc}$, we employ a kinematic fit
with the $\Delta E=0$ constraint. It is done by minimizing the 
$\chi^2$ function formed by the momenta of the daughter particles
\begin{equation}
  \chi^2=\sum\limits_i\frac{(p'_i-p_i)^2}{\sigma^2_{p_i}}\,,
\end{equation}
where $p_i$ and $\sigma_{p_i}$ are the measured momenta of the daughter 
particles and their errors obtained from the track fit, respectively, 
and $p'_i$ are the fitted momenta which satisfy the $\Delta E(p'_i)=0$
constraint. The use of $M_{\rm bc}$ constructed from the fitted momenta 
results in a certain improvement of its resolution and significantly 
reduces the dependence of measured mass on the absolute momentum 
calibration (see below). 

The precision of the momentum measurement has direct influence
on the $D$ mass measurement. The following sources of momentum
reconstruction uncertainties are considered in our analysis:
\begin{enumerate}
  \item {\bf Simulation of ionization losses in the detector material.}
        Reconstruction of cosmic tracks is used to check the validity
        of the simulation. We select the cosmic tracks that traverse 
        the vacuum chamber and fit their upper and lower parts separately. 
        The average difference of the upper and the lower track momenta due to 
        energy loss in the detector material is compared with the result 
        of the simulation.

  \item {\bf Absolute momentum calibration}
        (this is equivalent to the
        knowledge of the average magnetic field in the tracking system), 
        described by the scale
        coefficient $\alpha$ which relates the true track momentum
        $p_{\rm true}$ and the measured momentum $p$:
\begin{equation}
  p_{\rm true} = \alpha p\,.
\end{equation}
Then
\begin{equation}
    M_D=\sqrt{E^2_{\rm beam}-\alpha^2\left(\sum\limits_i \vec{p}_i\right)^2}\,, 
\label{mbc_eq}
\end{equation}
\begin{equation}
    \frac{dM_D}{d\alpha}\simeq -\frac{p_D^2}{M_D}\simeq -36\;
    \mbox{MeV}.
\label{dmda}
\end{equation}

The momentum scale can be calibrated using the same events as in
the $D$ mass measurement by measuring the average bias of the
$\Delta E$ value:
\begin{equation}
    \Delta E=\sum\limits_i \sqrt{M^2_i+\alpha^2 p^2_i}-E_{\rm beam}\,. 
\label{de_eq}
\end{equation}
Sensitivities to the scale coefficient $\alpha$ are given by
\begin{equation}
    \frac{d\Delta E}{d\alpha}\simeq
         \frac{p_K^2}{E_K}+\frac{p_{\pi}^2}{E_{\pi}}\simeq 1580\; \mbox{MeV}
\label{deda_kp}
\end{equation}
for $D^0\to K^-\pi^+$ decay, and 
\begin{equation}
    \frac{d\Delta E}{d\alpha}\simeq 1490\; \mbox{MeV}
\label{deda_kpp}
\end{equation}
for $D^+\to K^-\pi^+\pi^+$ decay. The numerical values of 
the $d\Delta E/d\alpha$
derivatives are obtained using the Monte Carlo (MC) simulation 
of the corresponding decays. 
When the kinematic fit with $\Delta E=0$ 
is employed for an $M_{\rm bc}$ calculation, such a correction 
is effectively 
applied to each event, and thus the dependence
of $M_D$ on the absolute momentum calibration is significantly reduced
(to $dM_D/d\alpha=-3$~MeV for $D^0\to K^-\pi^+$ and 
$-12$~MeV for $D^+\to K^-\pi^+\pi^+$).

As a cross-check, we also use other processes for the absolute momentum
calibration: the inclusive $K^0_S\to \pi^+\pi^-$ reconstruction and
$e^+e^-\to \psi(2S)\to J/\psi\pi^+\pi^-$ process. 

  \item {\bf Simulation of the momentum resolution.} Since the $D$ meson sample
        is limited, we use full MC simulation of the detector to determine
        the shapes of the signal distributions. The description of the
        momentum resolution in the simulation is adjusted using events
        of elastic $e^+e^-$ scattering, inclusive resonstruction of
        $K^0_S\to \pi^+\pi^-$ decay, and the process
        $e^+e^-\to \psi(2S)\to J/\psi\pi^+\pi^-$.
\end{enumerate}

\section{Analysis of $D^0\to K^-\pi^+$}

\label{kp_section}

The analysis uses a sample of 0.9 pb$^{-1}$ accumulated with the KEDR detector
at the energy of the $\psi(3770)$ resonance. Multihadron candidates which
contain at least three tracks close to the interaction region (transverse
distance from the beam $R<5$~mm, and longitudinal distance $|z|<120$ mm)
forming a common vertex are selected at the first stage of the analysis.
The pairs of oppositely charged tracks are taken as $D^0$ decay
candidates with the following requirements:
\begin{itemize}
  \item Number of track hits $N_{\rm hits}\ge 24$ 
        (the maximum number of hits per track is 48),
  \item Track fit quality $\chi^2/ndf<50$,
  \item Transverse momentum: $100$~MeV~$< p_T < 2000$~MeV.
  \item Energy of the associated cluster in the calorimeter
        $E<1000$ MeV.
\end{itemize}

We expect around 100 $D^0\to K^-\pi^+$ signal events for this sample.
In order to measure the $D^0$ mass most efficiently,
the unbinned maximum likelihood fit procedure
is used. Except for the $M_{\rm bc}$ variable, the likelihood function includes
two other variables which allow one to efficiently separate the signal from
the background: the energy difference $\Delta E=E_D-E_{\rm beam}$
(\ref{de_eq}) and the difference of the absolute values of momenta 
for $D$ decay products in the CM frame $\Delta|p|$.

The likelihood function has the form:
\begin{equation}
 -2\log\mathcal{L}({\bf\alpha})=-2\sum\limits_{i=0}^{N}\log p({\bf v}_i| {\bf\alpha})+
  2N\log\int\!\!p({\bf v}| {\bf\alpha}) d{\bf v}\,,
\end{equation}
where ${\bf v}=(M_{\rm bc}, \Delta E, \Delta |p|)$ are the variables that
characterize 
one event, $p({\bf v}| {\bf \alpha})$ is the probability
distribution function (PDF)
of these variables depending on the fit parameters
${\bf\alpha}=(M_D, \langle\Delta E\rangle, b_{uds}, b_{DD})$:
\begin{equation}
 p({\bf v}|{\bf\alpha}) = p_{sig}({\bf v}|M_D, \langle\Delta E\rangle)+
   b_{uds}p_{uds}({\bf v})+
   b_{DD}p_{DD}({\bf v})\,.
 \label{exp_pdf}
\end{equation}
Here $p_{sig}$ is the PDF of the signal events which depends on 
$M_D$ ($D^0$ mass) and $\langle\Delta E\rangle$ (the central value
of the $\Delta E$ distribution), $p_{uds}$ is the PDF for the
background process $e^+e^-\to q\overline{q}$ ($q=u,d,s$), and $p_{DD}$ is
the PDF for the background from $e^+e^-\to D\overline{D}$ decays
with $D$ decaying to all modes other than the signal one, $b_{uds}$ and $b_{DD}$ are
their relative magnitudes. The shape of the $p_{sig}$, $p_{uds}$ and $p_{DD}$
distributions is obtained from the MC simulation.
Such a fit procedure gives only a shape of the fitted distribution
without the absolute normalization. The numbers of signal and background events
can be extracted by taking the total number of events in a sample
and fractions of the corresponding events from the fit.

For a proper
calculation of $\Delta E=E_{\pi}+E_K-E_{\rm beam}$, the $\pi/K$
identification is needed. Presently it cannot be performed reliably in the
momentum range near 800 MeV. Fortunately, since the $D$ meson momentum
is small, the momenta of $K$ and $\pi$ differ by a small amount, and the
maxumum error (in the case of wrong mass assignment) is not larger than
30 MeV. Thus, we take the following combination as a $D$ meson energy:
\begin{equation}
  E'=(E_{K^-\pi^+}+E_{K^+\pi^-})/2\,,
  \label{eprime}
\end{equation}
where
\begin{equation}
  \begin{split}
    E_{K^-\pi^+} &= \sqrt{M_K^2+p_-^2}+\sqrt{M_{\pi}^2+p_+^2}\,, \\
    E_{K^+\pi^-} &= \sqrt{M_K^2+p_+^2}+\sqrt{M_{\pi}^2+p_-^2}\,.
  \end{split}
\end{equation}
The energy $E'$ calculated this way is practically unbiased
from the true energy $E$. A bias can appear if the detection efficiency
varies with momenta of the final state particles; we estimate the upper
limit of this bias to be 1.5 MeV. According to (\ref{dmda}) and 
(\ref{deda_kp}), this bias is propagated to an $M_{D^0}$ bias of 0.034 MeV.
$E'$ differs from $E$ by less than 15 MeV
in each event, this only slightly affects the $\Delta E$ error due to momentum
resolution.

\begin{figure}
  \centering
  \epsfig{file=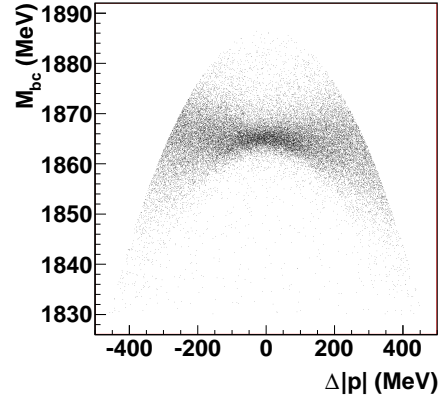,width=0.33\textwidth}
  \caption{Correlation of $M_{\rm bc}$ and $\Delta|p|$ variables for 
           $D^0\to K^-\pi^+$ decays (MC simulation).}
  \label{sig_mbcdp}
\end{figure}

Use of the $\Delta|p|$ variable allows us to obtain an estimate of the 
$M_{\rm bc}$ resolution on the event-by-event basis, thus improving the 
overall statistical accuracy of the measurement.
We use the fact that
this resolution depends strongly on decay kinematics ---
it can be up to three times better for events where the daughter
particles from $D^0$ decay move transversely to the direction of the $D^0$
($\Delta|p|$ is around zero for these events), than for events
where they move along this direction (see Fig.~\ref{sig_mbcdp}).

The variables $M_{\rm bc}$ and $\Delta|p|$ use the momenta of 
the daughter particles after the kinematic fit with the 
$\Delta E=0$ constraint, while $\Delta E$ is 
calculated using uncorrected momenta.
We select combinations that satisfy the following requirements for
the further analysis: $M_{\rm bc}>1700$~MeV,
$|\Delta E|<300$~MeV.

Simulation of signal events is performed with the MC generator
for $e^+e^-\to D\overline{D}$ decays where $D$-meson decays are simulated
by the {\tt JETSET 7.4} package~\cite{Sjostrand:1986hx},
and the radiative corrections are
taken into account in both initial (ISR, using the {\tt RADCOR}
package~\cite{radcor} with Kuraev-Fadin model~\cite{fadin-kuraev}), and
final states (FSR, the {\tt PHOTOS} package~\cite{photos}).
The ISR corrections use the
$e^+e^-\to D\overline{D}$ cross section dependence of the resonant
production of the $\psi(3770)$ according to a Breit-Wigner amplitude with
$M=3771$ MeV and $\Gamma=23$ MeV~\cite{pdg2006},
without the nonresonant contribution and taking into account
phase space dependence at the production threshold.
The full simulation of the KEDR detector is performed using the GEANT 3.21
package~\cite{GEANT}.

The PDF of the signal events $p_{sig}$ is a function of three parameters
$M_{\rm bc}$, $\Delta E$, and $\Delta|p|$. It is parameterized with 
the sum of two two-dimensional Gaussian distributions in
$M_{\rm bc}$ and $\Delta E$ (representing the core and the tails of 
the distribution) with a correlation and with the quadratic dependence of 
the $M_{\rm bc}$ resolution on $\Delta|p|$. The core
distribution is asymmetric in $M_{\rm bc}$ (with the resolutions
$\sigma_L(M_{\rm bc})$ and $\sigma_R(M_{\rm bc})$ for the left and right
slopes, respectively). The  $\Delta|p|$ distribution is
uniform with a small quadratic correction and with the kinematic constraint
$(\Delta|p|)^2<E^2_{\rm beam}-M^2_{\rm bc}$.
The parameters of the signal distribution are obtained from 
the fit to the simulated signal sample. 
The core resolutions obtained from the MC for $M_{\rm bc}$ are
$\sigma_L(M_{\rm bc})=0.98\pm 0.03$~MeV,
$\sigma_R(M_{\rm bc})=2.45\pm 0.06$~MeV (at $\Delta|p|=0$), the $M_{\rm bc}$
resolution at $\Delta|p|=200$ MeV is $4.6\pm 0.1$~MeV,
the core resolution of $\Delta E$ is $48.3\pm 0.3$~MeV.

The background from the continuum $e^+e^-\to q\bar{q}$
process (where $q=u,d,s$) is simulated using the {\tt JETSET 7.4}
$e^+e^-\to q\bar{q}$ generator. The PDF is parameterized as
\begin{equation}
\begin{split}
  p_{uds}(M_{\rm bc}, \Delta E, \Delta |p|) = &
    \exp\left(-k_1\left[1-\frac{M_{\rm bc}^2}{E_{\rm beam}^2}\right]-k_2\Delta
    E\right)\times\\
    &(1+k_3\Delta|p|^2)\,,
\end{split}
\end{equation}
where $k_i$ are  free parameters. The kinematic limit at
$M_{\rm bc}=E_{\rm beam}$ is provided by the 
$(\Delta|p|)^2<E^2_{\rm beam}-M^2_{\rm bc}$ constraint.

The background from $e^+e^-\to D\overline{D}$ decays is simulated using
the {\tt JETSET 7.4} generator, where the signal process $D^0\to K^-\pi^+$
is suppressed in the decay table. The PDF for $D\overline{D}$
background is parameterized with the function $p_{DD}$ of the same
form as for $p_{uds}$, with the addition of three two-dimensional
Gaussian distributions in $M_{\rm bc}$ and $\Delta E$. Two of them
describe the background from $D^0\to\pi^+\pi^-$
and $D^0\to K^+K^-$, while the third one is responsible for the
decays of $D$ mesons to three and more particles.

The combinatorial background coming from the signal events 
where one or more 
tracks were taken from the decay of the other $D$ meson, were studied 
using the signal MC sample. The distribution of fit variables for these 
events is similar to the background from the continuum events, and their 
fraction is 2.5\% of the number of signal events, which is negligible 
compared to the continuum contribution. We therefore do not treat this 
background separately, and its contribution is effectively taken by 
the continuum component.

\begin{figure*}[ht!]
  \epsfig{file=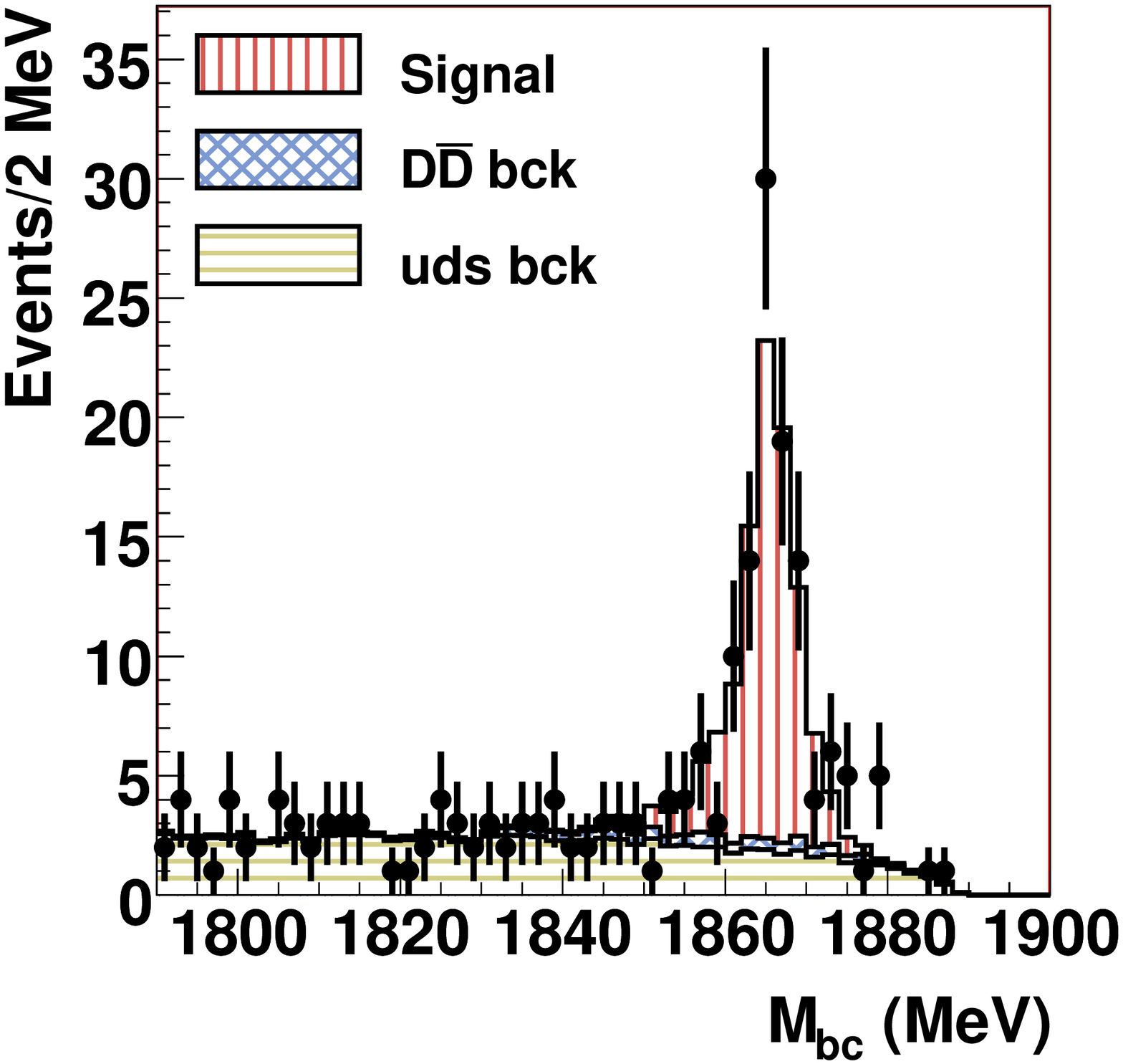,width=0.33\textwidth}
  \epsfig{file=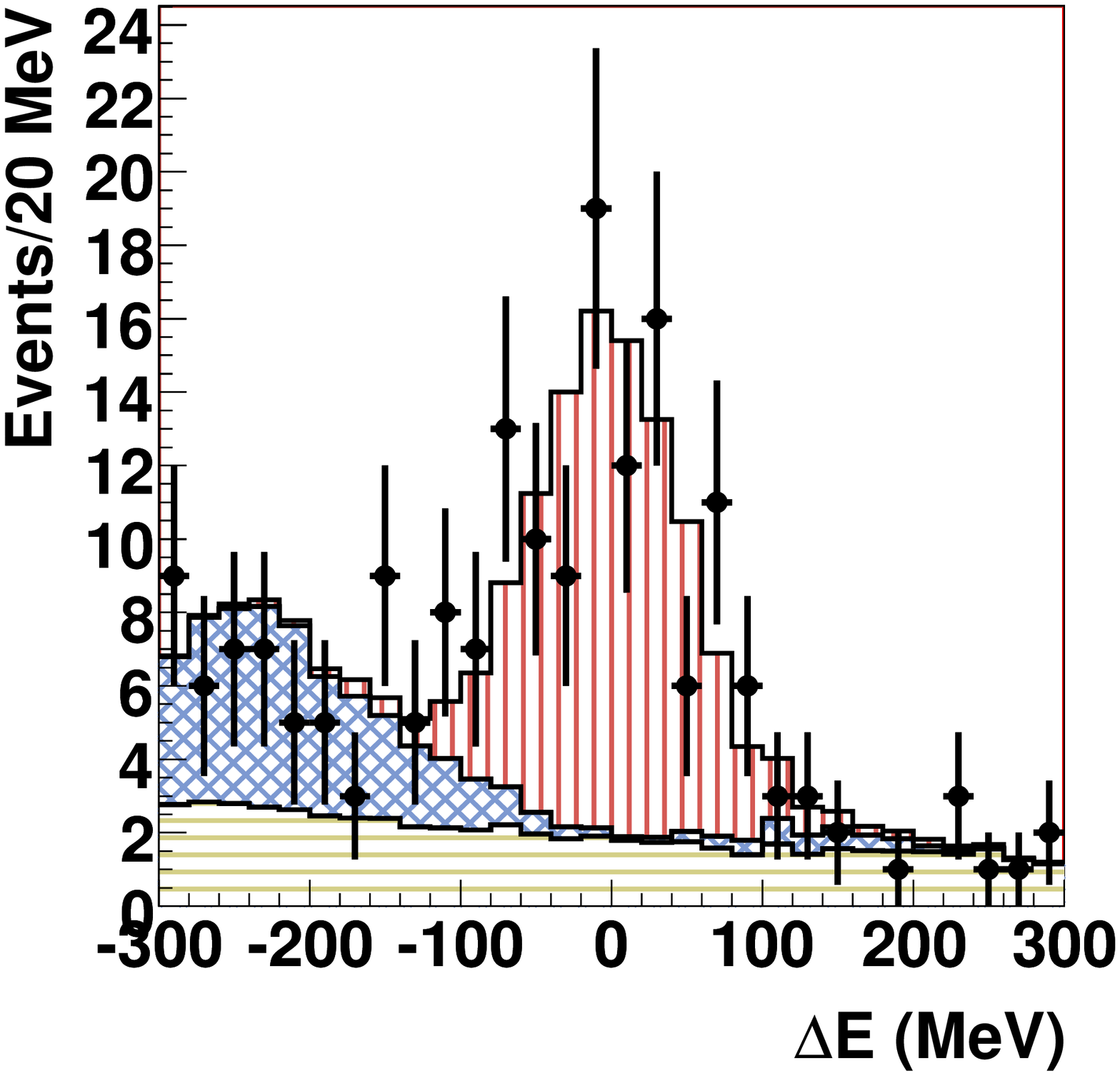,width=0.33\textwidth}
  \epsfig{file=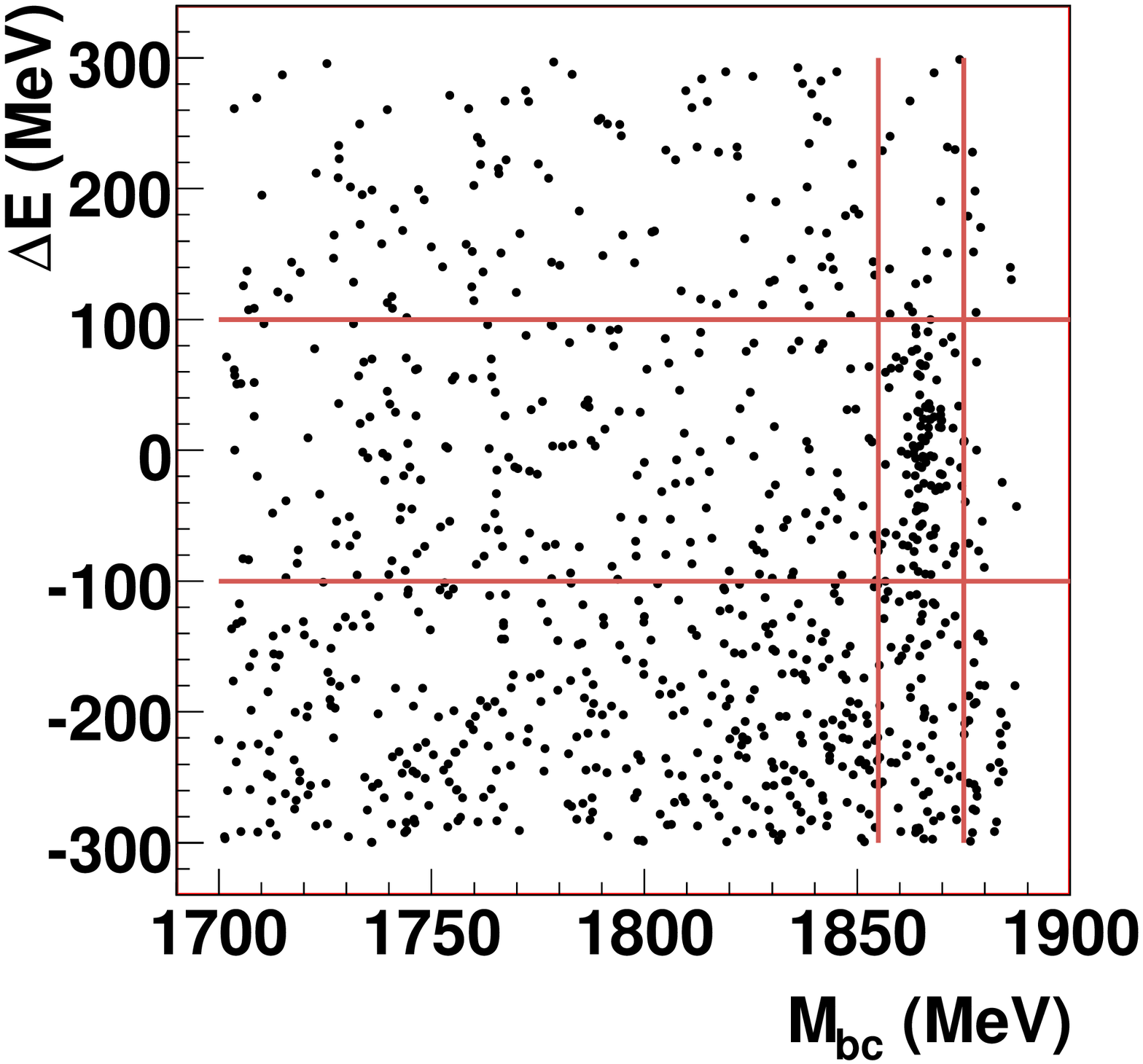,width=0.33\textwidth}
  \put(-380,140){\mbox{\bf(a)}}
  \put(-210,140){\mbox{\bf(b)}}
  \put(-34,140){\colorbox{white}{\bf(c)}}
  \caption{Experimental data (points with the error bars) and
           the results of the fit (histogram) for
           the $D^0\to K^-\pi^+$ decay.
           $M_{\rm bc}$ distribution for events with $|\Delta E|<100$~MeV~(a), 
           $\Delta E$ distribution for events with $1855$~MeV$<M_{\rm bc}<1875$~MeV~(b),            
           and the experimental $(M_{\rm bc}, \Delta E)$ scatter plot~(c). 
           }
  \label{kp_exp}
\end{figure*}

The result of the fit to the experimental data is shown in
Fig.~\ref{kp_exp}. In the fit we use the function (\ref{exp_pdf}) 
with $M_D$, $\langle\Delta E\rangle$
as well as the relative magnitudes of the continuum and $D\overline{D}$
backgrounds as free parameters.

\begin{table}
  \caption{Results of the fit to the $D^0\to K^-\pi^+$ data sample}
  \label{kp_results}
  \vspace{0.5\baselineskip}
  \centering
  \begin{tabular}{|l|l|}
    \hline
    $M_{D}$                          &            $1865.05\pm 0.33$ MeV \\
    $\langle\Delta E\rangle$         & \phantom{000}$-0.7\pm 7.3$ MeV \\
    Number of signal events          & \phantom{$-00$}$98.4\pm 13.1$ \\
    Number of $q\bar{q}$ events      & \phantom{$-00$}$18.3\pm 2.4$ \\
    Number of $D\overline{D}$ events & \phantom{$-000$}$4.8\pm 0.8$ \\
    \hline
  \end{tabular}
\end{table}

The momentum correction coefficient $\alpha$ is chosen to keep
the value of $\langle\Delta E\rangle$ close to zero. Event
selection is performed with $\alpha=1.030$; after 
the residual $\Delta E$ bias is taken into account 
its value is $\alpha=1.0304\pm 0.0046$.
The results of the fit are shown in Table~\ref{kp_results}.
The numbers of events are presented for the signal
region $|\Delta E|<100$~MeV, $1855$~MeV$<M_{\rm bc}<1875$~MeV.

To obtain the $D^0$ mass, one has to take into account a possible
deviation of the fit parameters $M_D$ and $\langle \Delta E\rangle$
from the true $D^0$ mass and energy.
In particular, the central value of $M_D$ can be shifted due to
the asymmetric resolution function and radiative corrections.
This deviation is corrected using the MC simulation. The final
value of the $D^0$ mass after the correction is
$M_{D^0}=1865.30\pm 0.33$ MeV.

\section{Analysis of $D^+\to K^-\pi^+\pi^+$}

\label{kpp_section}

The three-body decay $D^+\to K^-\pi^+\pi^+$ has more kinematic 
parameters and there is no simple variable (such as $\Delta|p|$ 
in the $D^0\to K^-\pi^+$ case), which determines the precision 
of the $M_{\rm bc}$ reconstruction. Therefore, we use only two 
variables, $M_{\rm bc}$ and $\Delta E$, in a fit of this mode. 

The mode $D^+\to K^-\pi^+\pi^+$ does not have a problem with $\pi/K$
identification for the $\Delta E$ calculation, since the sign of the kaon charge
is opposite to the pion charges and thus energies of all the particles
can be obtained unambiguously. The triplets of tracks with the charge of 
one of the tracks ("kaon") opposite to the charges of the two other 
tracks ("pions") are taken as $D^{\pm}$ decay candidates.

The requirements for the track selection are the same as in
the $D^0\to K^-\pi^+$
case. Since the significant part of the kaon tracks in the three-body decay
have relatively low momentum (under 500 MeV), an additional suppression
of the background from pions is possible using the TOF
system. The selection uses the following requirement on the flight time
for a kaon candidate, which hits the barrel part of the TOF system:
$\Delta T_{TOF} = T_{TOF}-T_K(p_K) > -0.8$ ns (or 2.3 times the 
flight time resolution), where
$T_K(p_K)$ is the expected flight time for a kaon with the
momentum $p_K$ and $T_{TOF}$ is the measured flight time.
As a result of  this requirement the background fraction 
is reduced by a factor of 2.3 for the continuum background and 3.3
for the $D\overline{D}$ 
background.


The $M_{\rm bc}$ variable uses the momenta of 
the daughter particles after the kinematic fit with the 
$\Delta E=0$ constraint. The variable $\Delta E$ is 
calculated using uncorrected momenta.
We select combinations that satisfy the following requirements for
the further analysis: $M_{\rm bc}>1700$ MeV,
$|\Delta E|<300$ MeV.

As in the case of $D^0\to K^-\pi^+$ decay, simulation is performed
using the $e^+e^-\to D\overline{D}$ generator taking into account the
ISR and FSR effects. 
The signal PDF $p_{sig}$ is parameterized in the same way 
as for the $D^0\to K^-\pi^+$ mode, but without $\Delta |p|$ dependence.
The core resolutions obtained from the MC for $M_{\rm bc}$ are
$\sigma_L(M_{\rm bc})=2.07\pm 0.05$~MeV,
$\sigma_R(M_{\rm bc})=2.52\pm 0.06$~MeV,
the core resolution of $\Delta E$ is $26.5\pm 0.4$~MeV.

To parameterize the continuum $e^+e^-\to q\bar{q}$ background,
we use the empirical function of $M_{\rm bc}$ proposed in the 
Argus experiment~\cite{argus} and the exponent of the quadratic form 
in $\Delta E$:
\begin{equation}
  \label{kpp_uds_param}
  p_{uds}(M_{\rm bc}, \Delta E) = y
    \exp\left(-k_1 y^2-[k_2+k_3 y^2]\Delta E + k_4\Delta E^2\right)\,,
\end{equation}
where $y=\sqrt{1-M^2_{\rm bc}/E^2_{\rm beam}}$. The coefficients $k_i$ are free
parameters in the fit. The coefficient $k_3$ is responsible for the $M_{\rm bc}$
dependence of the $\Delta E$ slope, which appears after
the kinematic fit to $\Delta E = 0$. The PDF for the
$e^+e^-\to D\overline{D}$ background $p_{DD}$ is parameterized with the 
distribution of the same form as for $p_{uds}$, with the addition of 
two two-dimensional Gaussian distributions in $M_{\rm bc}$ and $\Delta E$.
They describe the contributions of $D^+\to K^+K^-\pi^+$
and $D$ decays to four and more particles. The combinatorial background 
from the signal events as in the case of the $D^0\to K^-\pi^+$ mode is 
effectively taken into account by the continuum component.

\begin{figure*}
  \epsfig{file=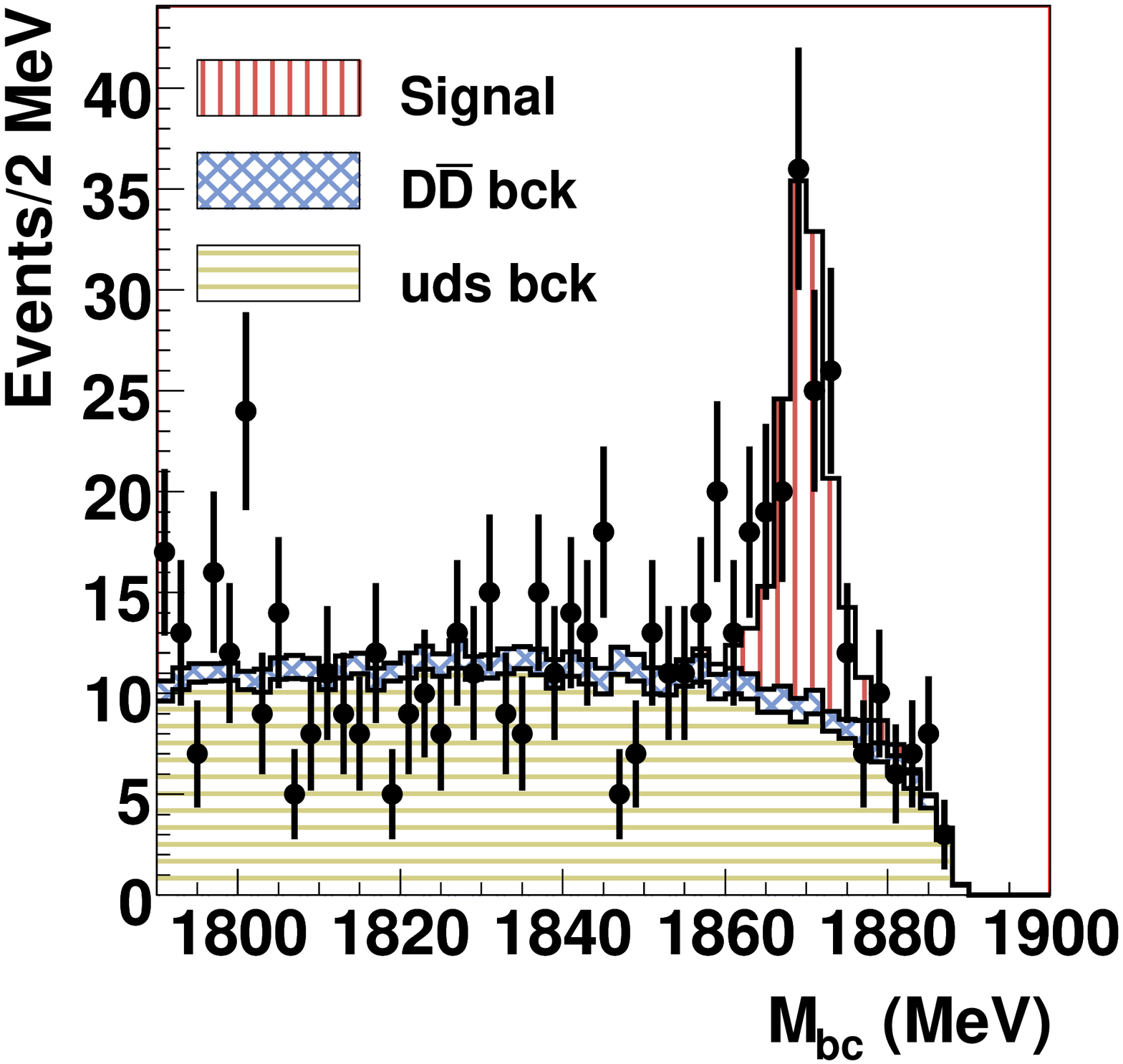,width=0.33\textwidth}
  \epsfig{file=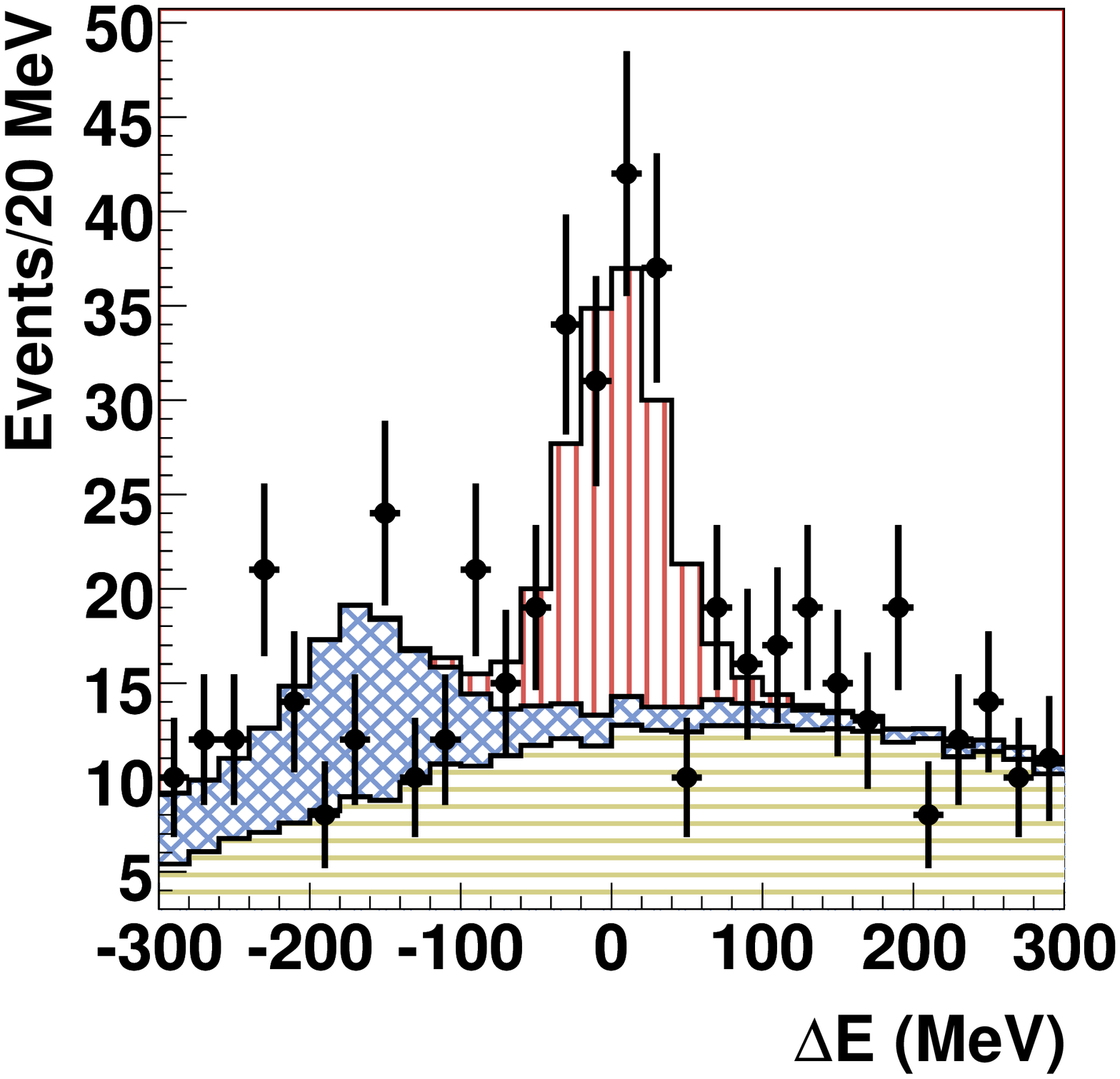,width=0.33\textwidth}
  \epsfig{file=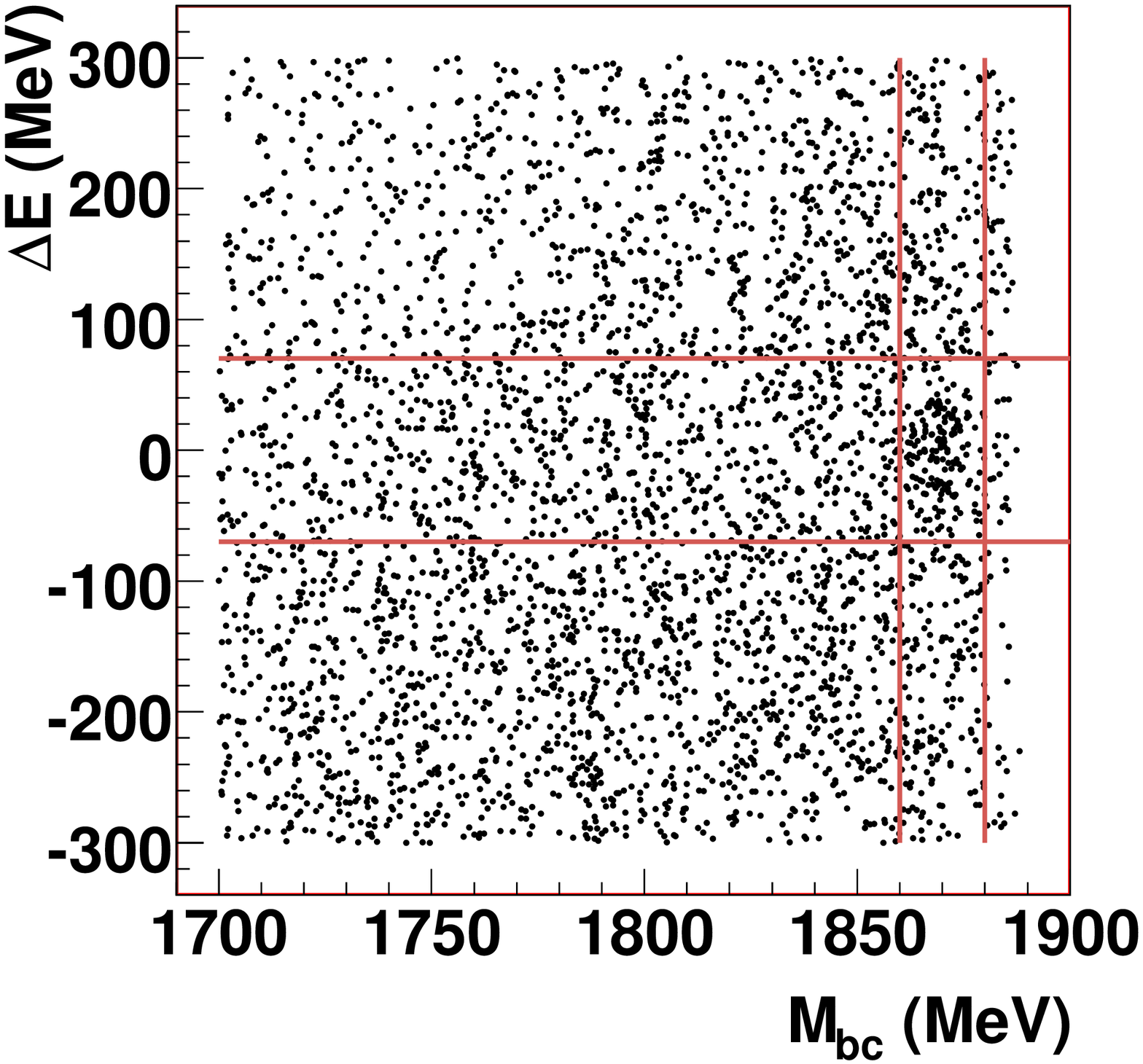,width=0.33\textwidth}
  \put(-380,140){\mbox{\bf(a)}}
  \put(-210,140){\mbox{\bf(b)}}
  \put(-35,140){\colorbox{white}{\bf(c)}}
  \caption{Experimental data (points with the error bars) and
           the results of the fit (histogram) for
           the $D^+\to K^-\pi^+\pi^+$ decay.
           $M_{\rm bc}$ distribution for events with $|\Delta E|<70$~MeV~(a), 
           $\Delta E$ distribution for events with $1860$~MeV$<M_{\rm bc}<1880$~MeV~(b), 
           and the experimental
           $(M_{\rm bc}, \Delta E)$ scatter plot (c).}
 \label{kpp_exp}
\end{figure*}

The result of the  fit to the data is shown in Fig.~\ref{kpp_exp}.
The momentum correction factor $\alpha$ is chosen such that
$\langle\Delta E\rangle$ is close to zero. The value $\alpha=1.027$
is used for event selection, and after taking into  account the residual
$\Delta E$ bias its value is $\alpha = 1.0252\pm 0.0035$. The results of
the fit are shown in Table~\ref{kpp_results}. The numbers of events are
shown for the signal region $|\Delta E|<70$ MeV, 
$1860$~MeV$<M_{\rm bc}<1880$ MeV.

\begin{table}
  \caption{Results of the fit to the $D^+\to K^-\pi^+\pi^+$ data sample}
  \label{kpp_results}
  \vspace{0.5\baselineskip}
  \centering
  \begin{tabular}{|l|l|}
    \hline
    $M_{D}$                          &           $1869.58\pm 0.49$ MeV \\
    $\langle\Delta E\rangle$         & \phantom{0000}$2.5\pm 5.0$ MeV \\
    Number of signal events          & \phantom{00}$109.8\pm 15.3$ \\
    Number of $q\bar{q}$ events      & \phantom{000}$85.3\pm 11.8$ \\
    Number of $D\overline{D}$ events & \phantom{000}$11.4\pm 2.2$ \\
    \hline
  \end{tabular}
\end{table}

As in the case of the $D^0\to K^-\pi^+$ mode, the $D^+$ mass obtained in the
fit is corrected for the bias of $M_D$ and $\Delta E$  using
MC simulation. The value of the $D^+$ mass after the correction is
$M_{D^+} = 1869.53\pm 0.49$ MeV.

\section{Systematic uncertainties}

\begin{table}
  \caption{Systematic uncertainties in the $D^0$ and $D^+$ mass measurements}
  \label{syst}
  \vspace{0.5\baselineskip}
  \centering
  \scalebox{0.92}{
  \begin{tabular}{|l|c|c|}
    \hline
                                             & $\Delta M_{D^0}$, MeV
                                             & $\Delta M_{D^+}$, MeV\\
    \hline
    Absolute momentum calibration            & 0.04 & 0.04 \\
    Ionization loss in material          & 0.01 & 0.03 \\
    Momentum resolution                      & 0.13 & 0.10 \\
    ISR corrections                          & 0.16 & 0.11 \\
    Signal PDF                               & 0.07 & 0.05 \\
    Continuum background PDF                 & 0.04 & 0.09 \\
    $D\overline{D}$ background PDF           & 0.03 & 0.06 \\
    Beam energy calibration                  & 0.01 & 0.01 \\
    \hline
    Total                                    & 0.23 & 0.20 \\
    \hline
  \end{tabular}
  }
\end{table}

The estimates of systematic uncertainties in the $D$ mass measurements
are shown in Table~\ref{syst}.

The contribution of absolute momentum calibration is determined by the
precision of the $\langle\Delta E\rangle$ measurement and is propagated
to the uncertainty of the mass measurement using the $dM_{\rm bc}/d\alpha$
dependence.
For the $D^0\to K^-\pi^+$ mode, the additional factor, which dominates the 
momentum calibration error, is a possible bias of the approximate 
$D$ energy calculation using Eq. (\ref{eprime}) 
in the absence of $\pi/K$ identification. However, due to smaller 
$dM_{\rm bc}/d\alpha$ value the momentum calibration uncertainty for 
this mode is close to the one for the $D^+\to K^-\pi^+\pi^+$ mode.

The uncertainty of the simulation of ionization losses in the detector
material is estimated by the variation of the corresponding correction
term within the limits given by the cosmic track measurement
($\pm 20$\%).

The uncertainty due to momentum resolution is estimated by using different
procedures matching the resolution in the simulation with the
experimental one (either by introducing a correction to the drift
curve of the DC, or by smearing the reconstructed momenta)
and by varying the tuning parameters responsible for the momentum
resolution matching within the limits given by the calibration processes.

The ISR correction uncertainty is dominated by the 
uncertainty of the energy dependence of the cross section 
$\sigma(e^+e^-\to D\overline{D})$. The default fit 
uses the $\psi(3770)$ parameters
from PDG-2006 for the cross section ($M=3771.1\pm 2.4$ MeV and
$\Gamma=23.0\pm 2.7$ MeV~\cite{pdg2006}). To estimate a systematic error,
these parameters are varied within their errors, also the PDG-2008
value is used ($\Gamma=27$ MeV~\cite{pdg2008}). In addition, the
nonresonant contribution is added incoherently to the 1 nb cross section
at the $\psi(3770)$ peak. The quadratic sum of deviations in $M$,
$\Gamma$ and non-resonant contribution is taken as the systematic error.
The model uncertainty of Kuraev-Fadin
formulae~\cite{fadin-kuraev} is small ($\sim 0.1\%$) and has a negligible 
effect on our results.

The uncertainty due to signal shape parameterization is estimated by
using the alternative shape with one Gaussian peak.

The continuum background shape uncertainty is estimated by using the
alternative generator for the system of pions with the varying
multiplicity in the simulation, and also by relaxing the background
shape parameters in the experimental fit.
The contribution of the $D\overline{D}$ background shape is estimated by
relaxing the relative magnitude of the Gaussian peaks and the
non-peaking component in the experimental fit,
and by excluding one of the Gaussian peaks from the background shape 
parameterization.
In the case of the $D^+\to K^-\pi^+\pi^+$ mode, the background shape
variation also includes the shapes obtained without a 
TOF requirement to take into account the uncertainty in the TOF simulation.

To check possible inconsistencies in the three-dimensional signal 
and background
description of the $D^0\to K^-\pi^+$ mode, we perform separate fits 
to data with different 
$\Delta|p|$ requirements. The results are consistent within statistical errors.

The error of the beam energy calibration is dominated by the precision
of the beam energy interpolation between successive energy measurements
using the resonant depolarization technique. It does not exceed 70~keV and
is of order 10~keV for most of the data sample. The uncertainty
due to beam energy calibration is estimated in the worst case of
a 100\% correlation between all energy measurements.

\begin{figure}[t]
 \epsfig{file=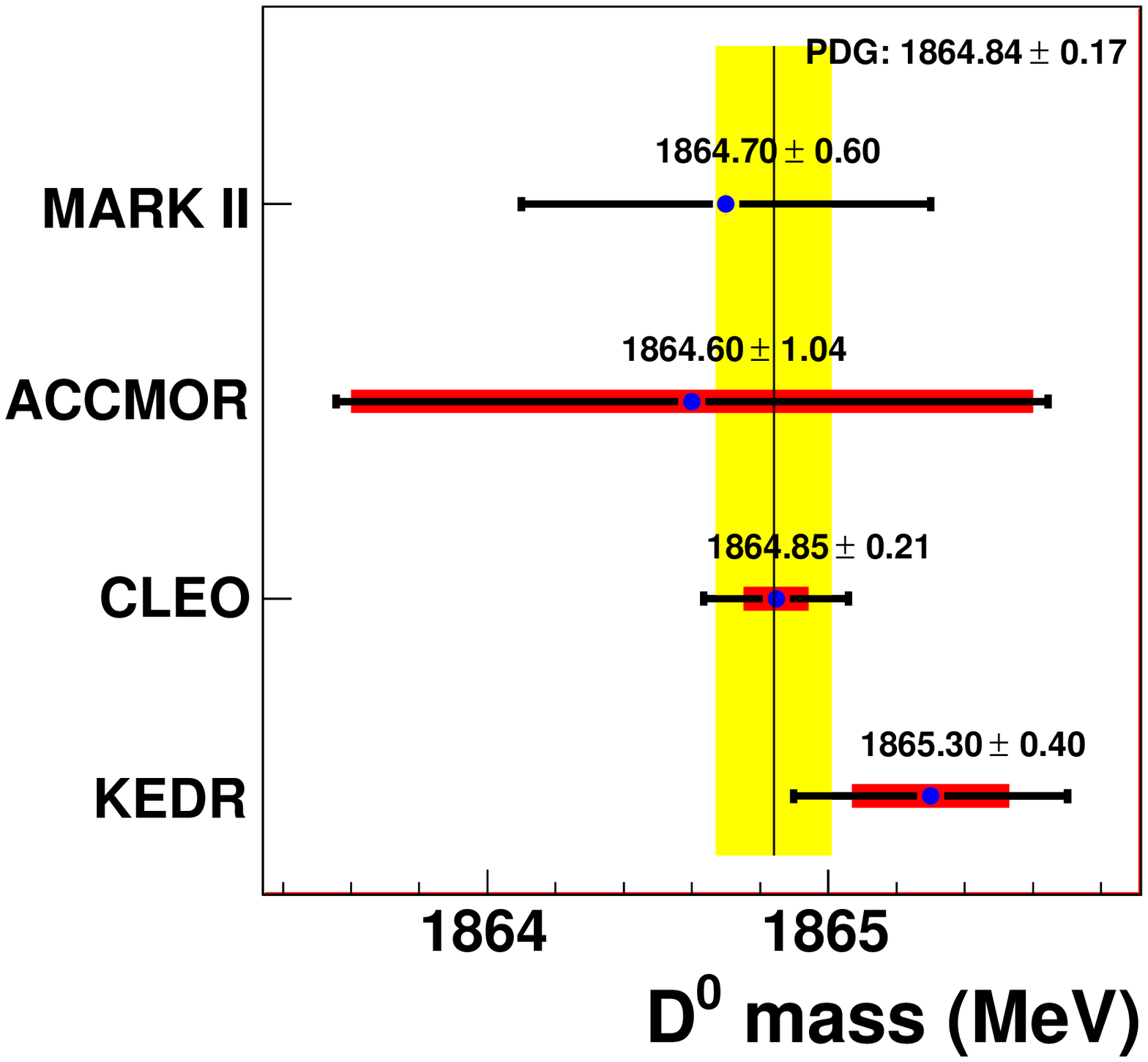,width=0.42\textwidth}
 \epsfig{file=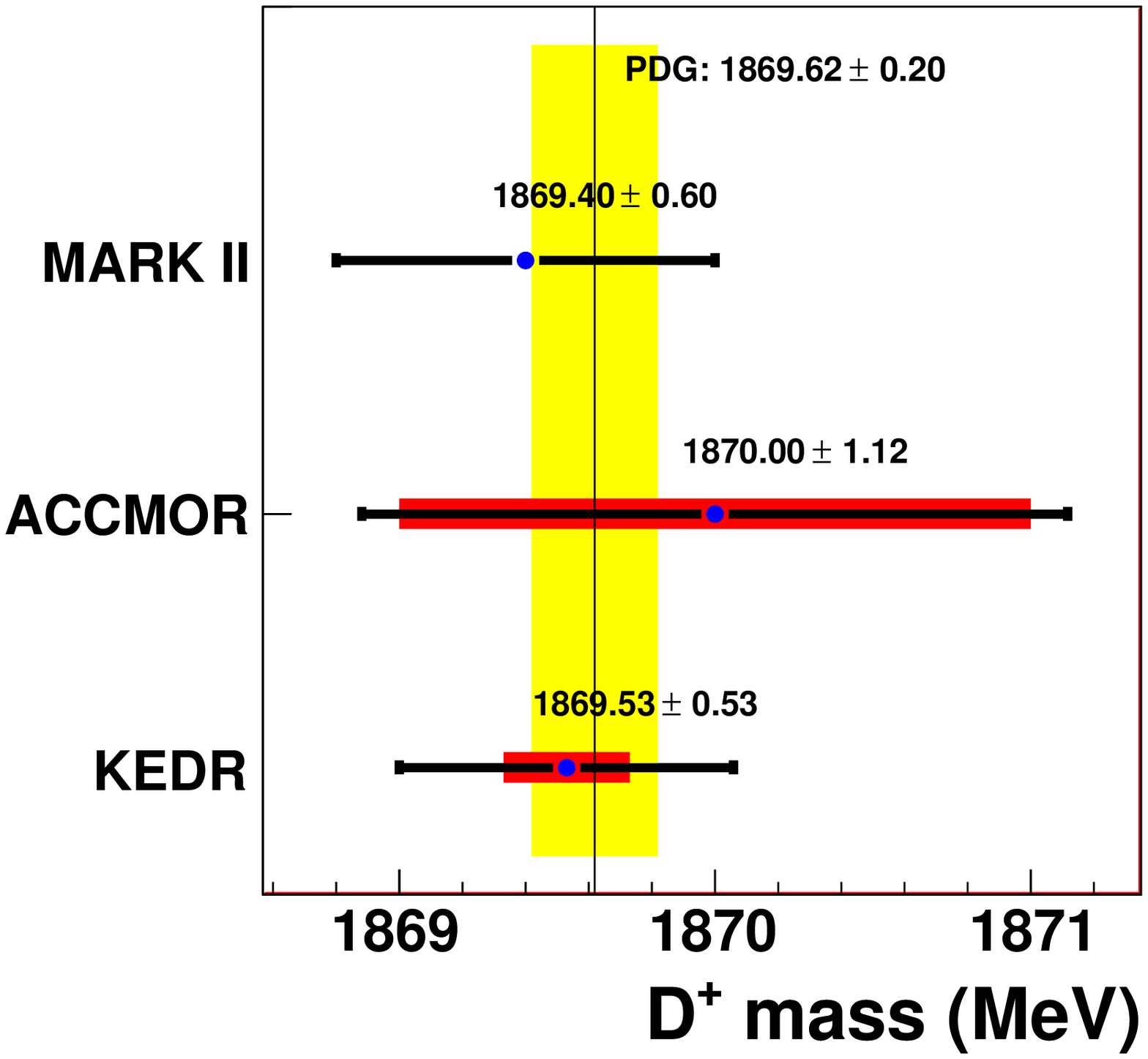,width=0.42\textwidth}
 \caption{Comparison of $D$ meson masses  with the
          other measurements. The thick and thin error bars show the
          systematic and the total errors, respectively. The shaded
          areas are the PDG-2008 values~\cite{pdg2008}.
          The PDG value for the $D^+$ is obtained using the measured
          mass difference of the $D^+$ and $D^0$ mesons. 
	  MARK-II does not quote the systematic error separately. }
 \label{dmass_comp}
\end{figure}

\section{Conclusion}

Masses of the neutral and charged $D$ mesons have been measured with the
KEDR detector at the VEPP-4M $e^+e^-$ collider operated in the region
of the $\psi(3770)$ meson. The analysis uses a data sample of
0.9 pb$^{-1}$ with $D$ mesons
reconstructed in the decays $D^0\to K^-\pi^+$ and $D^+\to K^-\pi^+\pi^+$. 
The values of the masses obtained are
\begin{itemize}
 \item $M_{D^0}=1865.30\pm 0.33\pm 0.23$ MeV,
 \item $M_{D^+}=1869.53\pm 0.49\pm 0.20$ MeV. 
\end{itemize}
The $D^0$ mass value is consistent with the more precise measurement of the
CLEO collaboration~\cite{cleo}, while that of the $D^+$ mass is presently
the most precise direct determination.

Comparison of the $D$ meson masses obtained in this analysis with the
other measurements is shown in Fig.~\ref{dmass_comp}.

\section{Acknowledgments}
We are grateful to the BINP administration for the permanent interest 
to this study and support of KEDR and VEPP-4M operation. This work is
partially funded by Russian Foundation for Basic Research, grants 
04-02-16712-a and 07-02-01162-a.

\bibliography{dmass}
\bibliographystyle{h-physrev}



\end{document}